.

# Increasing Robustness of the Anesthesia Process from Difference Patient's Delay Using a State-Space Model Predictive Controller

Saba Rezvanian[a,*], Farzad Towhidkhah[a], Nematollah Ghahramani[a], Alireza Rezvanian[b]

[a]Biomedical Engineering Department, Amirkabir University of Technology, Tehran, Iran
[b]Computer Engineering and Information Technology Department, Amirkabir University of Technology, Tehran, Iran

**Abstract**

The process of anesthesia is nonlinear with time delay and also there are some constraints which have to be considered in calculating administrative drug dosage. We present an Extended Kalman Filter (EKF) observer to estimate drug concentration in the patient's body and use this estimation in a state-space based Model of Predictive Controller (MPC) for controlling the depth of anesthesia. Bispectral Index (BIS) is used as a patient consciousness index and propofol as an anesthetic agent. Performance evaluations of the proposed controller, the results have been compared with those of a MPC controller. The results demonstrate that state-space MPC including the EKF estimator for controlling the anesthesia process can significantly increase the robustness in encountering patients' delay deviations in comparison with the MPC.

.

*Keywords:* Depth of Anesthesia (DOA); Patient's delay; PK-PD model; Model based Predictive Controller (MPC)

## 1. Introduction

Nowadays, anesthesia is one of the essential components of any surgical operation [1]. Anesthesiologists always want to be assured of the proper level of the patients' anesthesia. Automatic control methods are advantageous for this purpose which aims are administered the appropriate drug dosage to keep the patient on an adequate level of anesthesia. There should be an index which can be a good indicator of the depth of anesthesia. Among all the indexes which have been introduced in these years, BIS is one of the best [2] and we consider it as a deep index of anesthesia in this paper.

* Corresponding author. Tel.:+98-21-64545120.
E-mail address: s.rezvanian@aut.ac.ir

Recently, several methods have been introduced for controlling the Depth of Anesthesia (DOA) [3]. Fixed gain controllers such as P, PI, and PID strategies can perform well when used in clinical therapy and under certain conditions [4,5], on the other hand, It can lead poor performances because of the large variability between subjects and the delay which exists in the patient's model. The process of anesthesia is nonlinear with time delay and there are also some constraints which have to be considered in calculating administrative drug dosage. So, the MPC controller is a good choice for these kinds of systems. In [5] and [6], they used MPC for controlling the DOA. In [5], physiological model and also MAP are used as a sign of DOA. But, it seems that using other types of signals such as BIS or Auditory Evoked Potentials (AEP) are more valid for showing DOA level [7]. In [6] MPC with PK-PD model as a controller and BIS as hypnotic index are used. The results of this group were much better than those of [5], but since they used only the BIS as a feedback signal, their method did not have a good performance against disturbance and noise.

A lot of control methods described above have been used for controlling the DOA, but none of them pay attention to delay difference between patients. Whereas the delay of the patient is one of the individual patient's features and generally not obvious, the controller which endures larger difference delay, is more preferable. On the other hand, for controlling the DOA none of the methods has used state-space model. Since in the state feedback technique more information is used in feedback way, a designer has more freedom and the controller has a better performance. However, in this method all states may not be available. In our problem, there is no a direct access to the drug concentration in the patient's body, and therefore, the presence of an observer is essential.

In this paper, first, the EKF is used as a nonlinear observer in order to estimate the drug constraint in the patient's body and then, this observer is used in a state space based model predictive controller (MPC) for controlling the DOA which is evaluated by BIS. The proposed model is better comparison with that of the previous researches because it considers some individual parameters which will be introduced in section 2. In section 3 we explain the strategy and design of MPC with EKF algorithm. The result of the simulations is presented in section 4 and conclusions are reported in section 5.

## 2. Pharmacokinetic-Pharmacodynamic Model

In this section, we introduce a model of the BIS response to propofol infusion. The model is a series connection of two elements: a pharmacokinetic model and a Pharmacodynamic model. Pharmacokinetic models describe the dynamics of drug concentration in human body. We use a pharmacokinetic model based on the population of pharmacokinetic model was proposed by Schüttler and Ihmsen [8] because this model incorporates the individual patient's parameters.

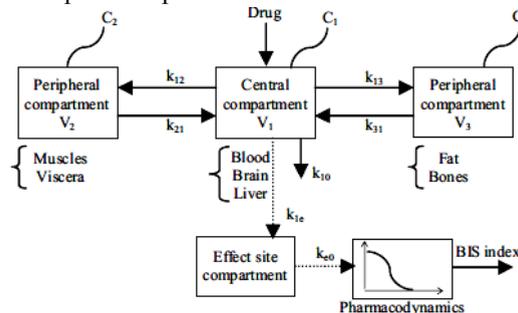

Fig. 1. Pharmacokinetics-Pharmacodynamic compartmental model with the effect site [9].

The original version of the Schüttler–Ihmsen model is given as equation (1). Here, $x_i$ is the concentration of propofol in compartments $i$; compartments *1*, *2*, and *3* correspond, respectively, to the central, shallow peripheral and deep peripheral compartments (Fig. 1). In addition u is the infusion rate of propofol, and $V_i$ are the clearance and volume of compartment $i$ respectively, given as functions of the patient's age and weight [6].

$$\frac{d}{dt}\begin{bmatrix} x_1(t) \\ x_2(t) \\ x_3(t) \end{bmatrix} = \begin{bmatrix} \frac{-(k_1+k_2+k_3)}{V_1} & \frac{k_2}{V_1} & \frac{k_3}{V_1} \\ \frac{k_2}{V_2} & -\frac{k_2}{V_2} & 0 \\ \frac{k_3}{V_3} & 0 & -\frac{k_3}{V_3} \end{bmatrix} \begin{bmatrix} x_1(t) \\ x_2(t) \\ x_3(t) \end{bmatrix} + \begin{bmatrix} \frac{1}{V_1} \\ 0 \\ 0 \end{bmatrix} . u(t) \quad (1)$$

An effect compartment is attached to the central compartment of the PK model to capture the complete response of the drug on specific endpoints (Fig. 1). It is assumed that the effect compartment receives a small mass of drug at a rate directly proportional to the central compartment drug concentration, which does not affect other time constants of the model. In steady-state, the concentration of this part can be related to the plasma concentration by the equation below:

$$\frac{dx_e}{dt} u(t) = k_e (x_1 - x_e) \quad (2)$$

Where $x_1$ and $x_e$ are the plasma concentrations and the concentration in the effect compartment, respectively [9].

A Pharmacodynamic model describes the relationship between the propofol concentration in the effect site compartment and the BIS. We use the sigmoid model $E_{max}$ because it is widely accepted in Pharmacodynamic studies of propofol. This model is given as

$$BIS(t) = BIS_0 \left(1 - \frac{x_e^\gamma}{x_e^\gamma + EC_{50}^\gamma}\right) \quad (3)$$

Where $BIS(t)$, $BIS_0$ are the current and awakening BIS, $EC_{50}$ is the concentration of the drug at which half of the maximum achievable effect is observed in the patient, $\gamma$ is the Hill coefficient [6]. According to [8], the PK model is obtained by knowing the age, weight of the patient, the sampling site and also the type of administration. For PD parameters we use the data gathered in [3] for 44 patients with different ages.

## 3. Depth of Anesthesia Control System

### 3.1. Extended Kalman Filter (EKF)

We cannot have access to the drug concentration in the patient's body; this problem can be solved by using an observer. An observer is integrated knowledge of plant and measurement dynamics, statistics of the process and measurement noise, and an initial condition, to yield an estimate of the state of the system. The well-known KF is an optimal estimator for linear systems [10]. In the sense of minimum error covariance, under the assumption that the plant and measurement dynamics are linear and known, the process and measurement noise are zero-mean Gaussian random processes and uncorrelated with each other, and the initial state of the system is a Gaussian random vector with known mean and covariance. The EKF is an extension of the original filter to nonlinear systems. Although the EKF has been a popular choice among researchers, the performance of the filter will depend upon the accuracy of the linear approximation. So, the linearization PD model around $EC_{50}$ was done at each stage of state approximation.

### 3.2. Model-Predictive Controller (MPC)

Manual control by anesthesiologist can be tedious, imprecise, time consuming and sometimes of poor quality, depending on his skills and judgment. Under dosing in a patient may cause pain and awareness during surgery, while overdosing may result in delayed recovery from anesthesia and may also result in respiratory and cardiovascular collapse. Closed-loop control may be improved the quality of drug administration, lessening the dependence of patient outcome on the skills of the anesthesiologist [7]. As anesthesia contains delay and there are also some constraints which should be considered on the infusion and the rate of drug infusion it seems that the MPC controllers belong to MPC is a good choice. The controller computes the vector of controls over prediction horizon by using optimization of a function:

$$J(N_1, N_2, N_u) = \sum_{j=N_1}^{N_2} \delta(j) \left[ y(t+j|t) - w(+j) \right]^2 + \sum_{j=1}^{N_u} \alpha(j) \left[ \Delta u(t+j-1) \right]^2 \quad (4)$$

Where $N_1$ and $N_2$ are the minimum and maximum costing horizon, $N_u$ is the control horizon, $w$ is the future set-point, $\alpha(j)$ and $\delta(j)$ are the weighting sequences. The aim of predictive control is to keep the output as near as $w$, for this purpose (5) should be minimized by considering the constraints defined in (5). For solving equation (4) subject to the following constraints, MATLAB quadratic program is used.

$$-0.2 \leq \Delta u(t+j-1) \leq 0.2; \qquad 0 \leq u(t+j-1) \leq 300 \frac{\mu g}{kg.min} \qquad (5)$$

Considering these constraints in controlling the process is necessary because first of all, the drug input cannot be negative and on the other hand the maximum effect of the drugs is defined since the administration of a drug more than a special amount is useless [9].

## 4. Simulation Results

The PK-PD model was used for designing the controller. The patient's specifications used in these simulations extracted from [3] and are given in Table 1.

Table 1. Patients' specifications [3]

| Parameter | $V_1$ (L) | $k_{10}$ (s$^{-1}$) | $k_{21}$ (s$^{-1}$) | $k_{12}$ (s$^{-1}$) | $k_{31}$ (s$^{-1}$) | $k_{13}$ (s$^{-1}$) | $k_{e0}$ (s$^{-1}$) | $T_d$ (s) | $BIS_0$ | $\gamma$ | $EC_{50}$ (µg/mL) |
|---|---|---|---|---|---|---|---|---|---|---|---|
| Nominal | 9.5855 | 0.0028 | 8.495e-4 | 0.0042 | 6.182e-5 | 0.0017 | 39e-3 | 12.9 | 100 | 2 | 3.3 |
| Patient 1 | 10.450 | 0.0029 | 8.506e-4 | 0.0044 | 6.659e-5 | 0.0018 | 24.8 e-3 | 4 | 100 | 2 | 2.7 |
| Patient 2 | 8.947 | 0.0027 | 8.485e-4 | 0.0042 | 5.810e-5 | 0.0017 | 83.1 e-3 | 29 | 100 | 2.3 | 4 |

A nominal model was considered by using average data and the design of controllers (state-space MPC couple with EKF and MPC) have been done based on this nominal patient model. The simulation results of both controllers are shown in Fig. 2 and Fig. 3.

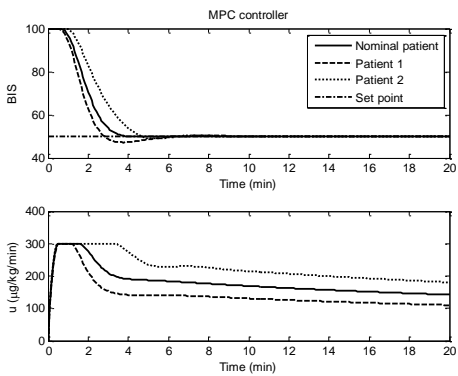
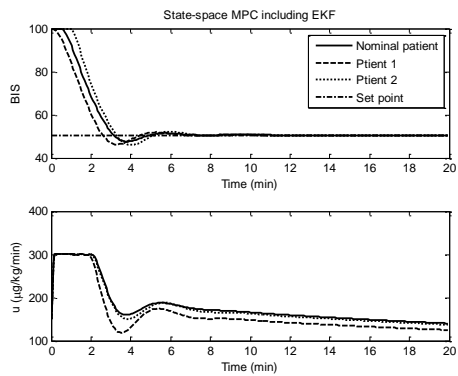

Fig. 2. Simulation results of the MPC controller. The upper panel shows the BIS, and the lower panel shows the infusion rate of the propofol for the nominal model (solid line), Patient 1 (dashed line) and Patient 2 (dotted line).

Fig. 3. Simulation results of the state-space MPC including the EKF. The upper panel shows the BIS, and the lower panel shows the infusion rate of the propofol for the nominal model (solid line), Patient 1 (dashed line) and Patient 2 (dotted line).

The simulations show, the methods have an acceptable performance. The amount of administered drug in purposed method is less than that of MPC controller, although the MPC controller shows less undershoots. If there is difference between the patient delay and nominal delay (Fig. 4 and Fig. 5) purposed method has better performance.

Table 2. Maximum range of tolerable delay for controllers

| Controller | Tolerable delay (Sec) |
|---|---|
| MPC | 30 |
| State-space MPC including EKF | 45 |

While the difference of delay is less than values in Table 2, the settling time slightly increases, but BIS stays in the acceptable bound. As mentioned in Table 2, purposed method can cover more delay difference than MPC controller.

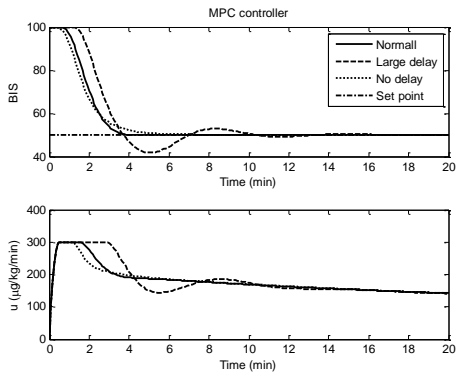

Fig. 4. Closed loop response of the anesthesia process in the nominal condition (solid line) and in presence of maximum tolerable delay (dashed line) and no delay (dotted line) for MPC controller.

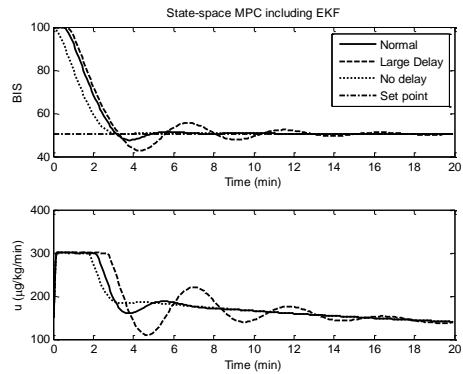

Fig. 5. Closed loop response of the anesthesia process in the nominal condition (solid line) and in presence of maximum tolerable delay (dashed line) and no delay (dotted line) for the state-space MPC including the EKF.

## 5. Conclusion

In this paper, the DOA was controlled by a model predictive controller. In this system the BIS was considered as the index of consciousness and propofol was considered as an anesthetic drug. We used EKF as an observer for the estimation of drug in the patient's body. When there isn't any uncertainty, simulations have shown that purposed method have similar performance with MPC. The results showed that the state-space MPC including the EKF estimator for controlling the anesthesia process can significantly increase the robustness in encountering patients' delay deviations in comparison with the MPC. Whereas the delay of the patient is one of the individual patient's feature and generally not obvious, the controller which endures larger difference from nominal case, is more preferable.